\documentclass[final,5p,times,twocolumn]{elsarticle}

\usepackage{amssymb}
 \biboptions{sort&compress}

\usepackage{float}
\usepackage{subfig}
\usepackage[export]{adjustbox}
\usepackage{placeins}
\usepackage{booktabs}
\newcommand{\head}[1]{\textnormal{\textbf{#1}}}
\usepackage{amsmath,bm}
\usepackage{pdflscape}
\usepackage{url}
\usepackage{textgreek}
\usepackage[final]{microtype}
\usepackage{multirow}

\usepackage[usenames]{color}

\usepackage{adjustbox}

\usepackage{dblfloatfix}

\newcommand{\beginsupplement}{%
        \setcounter{table}{0}
        \renewcommand{\thetable}{S\arabic{table}}%
        \setcounter{figure}{0}
        \renewcommand{\thefigure}{S\arabic{figure}}%
        \setcounter{section}{0}
        \renewcommand{\thesection}{S\arabic{section}}
     }

\newenvironment{itemize*}%
  {\begin{itemize}%
    \setlength{\itemsep}{0pt}%
    \setlength{\parskip}{0pt}}%
  {\end{itemize}}
  
\newenvironment{enumerate*}%
  {\begin{enumerate}%
    \setlength{\itemsep}{0pt}%
    \setlength{\parskip}{0pt}}%
  {\end{enumerate}}

\makeatletter
\renewcommand\paragraph{%
   \@startsection{paragraph}{4}{0mm}%
      {-\baselineskip}%
      {.5\baselineskip}%
      {\itshape\normalsize}}
\makeatother

\journal{Additive Manufacturing}

\makeatletter
\def\ps@pprintTitle{%
 \let\@oddhead\@empty
 \let\@evenhead\@empty
 \def\@oddfoot{}%
 \let\@evenfoot\@oddfoot}
\makeatother

\begin{document}

\begin{frontmatter}

%\title{High-Fidelity Optical Process Monitoring for Laser Powder Bed Fusion via Aperture Division Multiplexing}

 \title{High-Fidelity Optical Monitoring of Laser Powder Bed Fusion via Aperture Division Multiplexing}

\author[MIT]{Ryan~W.~Penny\corref{cor1}}
\ead{rpenny@mit.edu}
\author[MIT]{Zachery~Kutschke}
\author[MIT]{A.~John~Hart\corref{cor1}}
\ead{ajhart@mit.edu}

\address[MIT]{Department of Mechanical Engineering, Massachusetts Institute of Technology, 77 Massachusetts Avenue, Cambridge, 02139, MA, USA}
%\address[TUM]{Institute for Computational Mechanics, Technical University of Munich, Boltzmannstra{\ss}e 15, Garching b. M{\"u}nchen, Germany}

\cortext[cor1]{Corresponding author}

\begin{abstract}

Qualification of high-performance metal components produced by laser powder bed fusion (LPBF) must identify process-induced porous defects that reduce ductility and nucleate fatigue cracking.  Detecting such defects via optical monitoring of LPBF provides a path towards in-process quality control without downstream testing such as by computed tomography.  However, integration of in-process sensing with LPBF is hampered by geometric and optical complications and, as a result, it has yet to be proven that the finest pores that limit component fatigue life can be resolved via \textit{in situ} data.  We present aperture division multiplexing (ADM) as a method for simultaneously focusing the process laser and providing unobstructed optical access for high-fidelity process monitoring using a common optic.  Construction of an ADM optic of achieving imaging at $50$~\textmu m spatial resolution in the mid-wave infrared is described, and this optic is demonstrated on a production-representative LPBF testbed.  High-speed infrared video data are correlated to micro-CT measurement of pores as fine as $4.3$~\textmu m, through multiple process signatures, establishing the promise of ADM for qualification of LPBF component fatigue performance.

\end{abstract}

\begin{keyword}
additive manufacturing \sep process monitoring \sep quality control \sep porosity \sep infrared
\end{keyword}

\end{frontmatter}

\section{Introduction}

While laser powder bed fusion (LPBF) is the foremost technology for metal additive manufacturing~\cite{Wohlers2024}, efficiently qualifying the mechanical performance of components fabricated by LPBF remains an open challenge~\cite{Everton2016, Grasso2016, McCann2021, Chua2024}.  Porous defects are highly characteristic of LPBF, and achieving full density is complicated by a narrow range of suitable process parameters and considerable stochastic variation in feedstock delivery~\cite{Seifi2016, Zhang2017, Duplessis2019, Rombouts2005, Zhang2017, Meier2017}.  Typical density of carefully fabricated LPBF components is greater than $99$\%; however, even at this level, pores cause an outsized reduction in component fatigue life~\cite{Carlton2016, Nadot2019, Yadollahi2018, Poulin2018}.  Post-process identification of these internal defects using legacy nondestructive testing techniques is sharply limited by cost, time, resolution, and assessment domain (part size)~\cite{Chua2024}.  As a result, the inability to precisely bound component life impedes application of LPBF to cyclically-loaded aerospace~\cite{Blakey2021, Gruber2023} and automotive components~\cite{Nadot2019}, and is even implicated in premature failure of Ti-6Al-4V orthopedic implants~\cite{Lam2022}.  In-situ process monitoring is widely investigated as a potential solution, although improved sensitivity is requisite for reliably detecting the finest pores that affect fatigue performance with this qualification strategy~\cite{Everton2016, Grasso2016, McCann2021, Chua2024}.

Pores in LPBF components are often caused by improper print parameters including laser power and scan speed.  Lack-of-fusion (LoF) porosity occurs if insufficient energy is applied to fully melt the irradiated material~\cite{Thomas2016, Kasperovich2016, Zhang2017}. LoF pores are commonly characterized by irregularly shaped void regions around entrapped powder particles.  Keyhole porosity arises at the opposite, high-energy-density regime, wherein recoil pressure exerted by evaporated material causes a deep depression in the meltpool surface~\cite{King2014}; the shape of the fluid surface is unstable and results in rapidly varying interaction with process laser energy~\cite{Ye2019, Kouraytem2019}. This instability can nucleate porosity~\cite{King2014, Calta2018}, especially at abrupt changes in laser scan direction~\cite{Calta2019}.  The space of power and scan speed combinations is also bounded by meltpool length, even at otherwise acceptable energy density, where long meltpools break into large beads due to Rayleigh-Taylor instability~\cite{Chivel2013}.  This defect is known as balling and causes porosity by disturbing powder spreading and material consolidation as subsequent layers are fused~\cite{Chivel2013}.  A final source of porosity is gas entrapment.  One mechanism of gas porosity is due to gas that becomes dissolved in molten material; the gas comes out of solution to form bubbles as the metal cools and voids are generated where these bubbles are unable to reach the melt pool surface prior to solidification~\cite{Zhang2017}.   Additionally, porosity can remain from bubbles that are inadvertently frozen into the powdered feedstock at its time of manufacture, and may not escape the melt pool due to the high cooling rates and strong convective fluid flows~\cite{Cunningham2017, Tammas2016}.

\begin{figure}[htbp]
\begin{center}
	{
	\includegraphics[trim = {0.2in 3.75in 1.4in 1in}, clip, scale=.5, keepaspectratio=true]{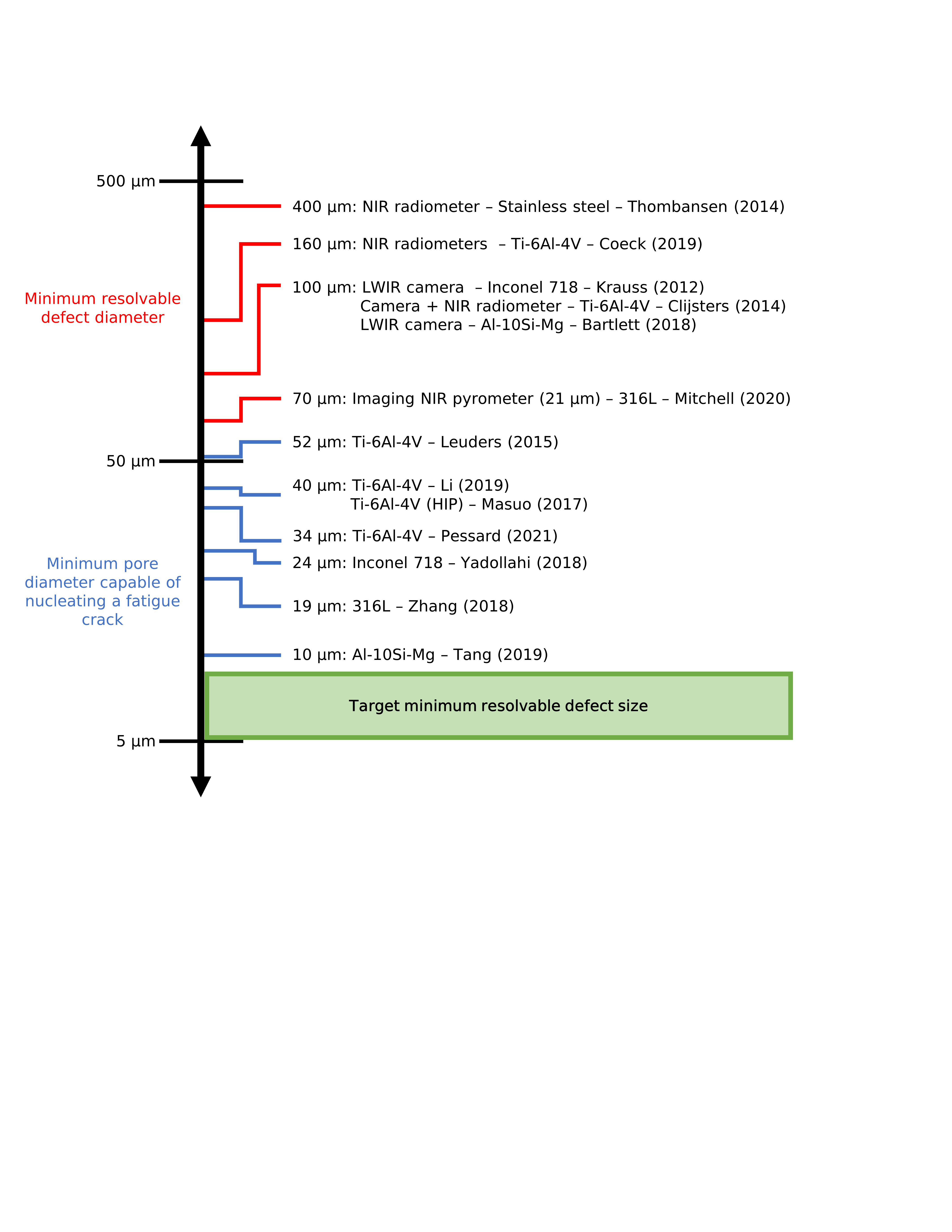}
	}
\end{center}
\caption{Comparison of minimum detectable pore size reported for various in-situ monitoring techniques, compared to the minimum pore diameter capable of nucleating a fatigue crack, for common LPBF materials. Data compiled from~\cite{Thombansen2014Tracking, Coek2019, Krauss2012, Clijsters2014, Bartlett2018, Mitchell2020, Leuders2015, Li2019, Masuo2017, Yadollahi2018, Zhang2018, Tang2019, Pessard2021}.}
\label{fig:IntroNumberLine}
\end{figure}

Pores have a deleterious effect on component mechanical properties~\cite{Carlton2016, Nadot2019, Leuders2013,Mercelis2006, Bartlet2019, Bhandari2022,Ronneburg2020, Jost2021, Fiocchi2021, Salarian2020}, and can reduce mechanical strength and ductility, act as stress concentrations, and nucleate fatigue cracks~\cite{Carlton2016, Nadot2019, Leuders2013, Bhandari2022}. Several studies more narrowly contemplate how pore size affects component fatigue life~\cite{Leuders2015, Li2019, Masuo2017, Yadollahi2018, Zhang2018, Tang2019, Pessard2021}.  While sensitive to material and loading conditions, reported values of the minimum pore size necessary to nucleate a fatigue crack are provided in Fig~\ref{fig:IntroNumberLine}.  For example, minimum pore diameters for Ti-6Al-4V are found to span $34$ to $52$~\textmu m~\cite{Leuders2015, Masuo2017, Li2019, Pessard2021}.  This is not to say that a component with finer pores will not fail due to fatigue, or even that a fine pore will never contribute to failure, but rather that the rate of damage nucleation from these sufficiently fine pores is comparable to that predictable from geometric and microstructural features.

It therefore follows that certifying the absence of pores of sizes greater than a characteristic dimension enables more precise bounding of the fatigue life of LPBF components.  Computed tomography (CT) is presently the leading technology for internal defect detection; however, as explained by du Plessis and coworkers, CT has difficulty resolving features on this size scale~\cite{DuPlessis2018}.  Specifically, they recommend that the voxel size in the reconstruction should be a third of the smallest pore size one must resolve and that the minimum voxel size is roughly $1/2000$\textsuperscript{th} the largest characteristic dimension of the component.  Linking these heuristics using linear dimensions, a voxel size of about $3.3$~\textmu m is necessary to resolve pores that are about $10$~\textmu m in diameter, and this is only possible if the part is smaller than $6.6$~mm.  As the build volume of large LPBF machines can be $600\times 600\times 600$~mm$^3$~\cite{Nikon2024} or larger, an alternative solution is necessary to resolve this problem at the industrial scale.

\subsection{Optical Monitoring of LPBF}

Optical process monitoring is often applied to detect component defects, or, equivalently, to certify their absence in LPBF~\cite{Spears2016, Grasso2021, Everton2016}.  Approaches to this task may be sorted into categories by three key attributes.  First is the dimensionality of the sensor, or, effectively, whether a point measurement is made (e.g., with a photodiode) or an area is spatially resolved (i.e., a camera is used).  Second, the field of view of the sensor remains stationary in some implementations, and in others it is scanned along with the process laser.  Third, often driven by the sample rate of the sensor, is the time scale of the process signatures extracted from the sensor data.  Here, defects may be detected from fast, transient features that indicate instability of the fusion process (e.g., at the time scale the laser spot traverses a point on the build surface), or longer duration signatures such as cooling rate.  An effort is made in the following text to accordingly classify and compare the most consequential prior art in the field.  Figure~\ref{fig:IntroNumberLine} additionally plots estimated detectable pore size where available.  It makes the present challenge obvious: none of the optical process monitoring techniques described are capable of resolving the finest pores that can influence fatigue life of LPBF components.

\subsubsection{Off-Axis Camera Techniques}

Using the aforementioned taxonomy, a first category of monitoring techniques uses a stationary camera that observes the LPBF process at an angle (off axis) to the nominal (vertical) path of the process laser~\cite{Zhirnov2015, Yang2020, Scime2019Monitoring, Le2021Monitoring, Lane2016, Doubenskaia2015, Williams2019, Bruna2018, Estalaki2022}.  Early work in this area by Krauss and coworkers describes a LWIR (long-wave infrared) microbolometer array with a $50$~Hz frame rate, which they use to detect flaws $100$~\textmu m and greater from characteristics of radiance profiles~\cite{Krauss2012}.  Later work by Krauss uses thermal diffusivity and peak temperature as process signatures from the fabrication of stainless steel cubes~\cite{Krauss2014}.  Cubes that delaminate from the substrate or exhibit high porosity or show substantial reductions in thermal diffusivity and corresponding increases in peak temperature.  Further, porosity explains $70$\% of the variance in optically measured thermal diffusivity across samples.  A similar instrument is described by Bartlett and colleagues, who study fusion of Al-10Si-Mg with a slow ($7$~FPS) LWIR camera. They report the ability to detect 82\% of lack-of-fusion defects with a diameter of $100$~\textmu m or greater, but only detect one third of equivalently sized keyhole pores~\cite{Bartlett2018}.  

Likewise, many studies investigate process monitoring in the MWIR (mid-wave infrared).  Mohr et al. consider time above a $700$~K threshold as a process signature and, while achieving qualitatively promising results, particularly highlight that the location of a component flaw is not necessarily coincident with the location of the process signature deviation that it manifests~\cite{Mohr2020}.  Foster and coworkers apply the same approach to relate time-temperature history of Inconel 718 specimens to attributes of metallurgical microstructure, but also report some ability to detect build failure~\cite{Foster2018}.  Meltpool length and cooling rate are extracted from high-speed video data in~\cite{Heigel2020}, showing that these signatures change when printing bridge-like features.  

Finally, work by Lough and colleagues monitors LPBF in the SWIR (short-wave infrared) using time above threshold and maximum radiance as process signatures, and establish positive correlations to component attributes including microhardness and porosity~\cite{Lough2020}.  The instrument is calibrated to temperature in a follow-up report, where a rigorous statistical approach is used to quantify the performance of time above threshold and maximum temperature in predicting keyholing and lack-of-fusion porosity~\cite{Lough2022}.

Off-axis imaging is also used in spectrally-resolved imaging modalities.  Mitchell and colleagues use an imaging two-color pyrometer wherein two visible high speed cameras are equipped with narrow bandwidth filters; the authors report the ability to reliably detect $70$~\textmu m diameter pores from temperature data collected therewith~\cite{Mitchell2020}.  A similar strategy is employed by Furumoto~\cite{Furumoto_2022}, who use different channels of a high speed camera with a color sensor, and this idea is further investigated in~\cite{Myers2023}.

\subsubsection{On-Axis Camera Techniques}

Cameras are also used on-axis in LPBF, where a dichroic mirror is used to combine the laser path and monitoring path, upstream of the galvanometer mirrors.  Accordingly, the field of view of the camera is scanned along with the laser spot.  An instrument comprising on-axis visible and NIR (near infrared) cameras is described in~\cite{Demir2018} and, using the summed values of the visible sensor, correlations are established to applied laser energy and to component porosities greater than 1\%.  A comparable limit is reported by deWinton et al. when using a $100$~kHz visible camera~\cite{Dewinton2021}.  They extract process signatures including meltpool maximum and average radiance in counts, as well as a rough estimate of a photodiode signal by summing pixels proximal to the meltpool; however, the authors conclude that none of these process signatures can qualify a part to better than $0.5$\% porosity.  Microstructure evolution of LPBF Ti-6Al-4V is studied with an on-axis CCD camera in~\cite{Yadroitsev2014}, where pixel intensities are calibrated to temperature and used to assess melt pool dimensions in turn.  In another area of advancement, Vasileska and coworkers use a high speed, on-axis visible camera to measure meltpool size, and apply the resulting data for layerwise feedforward control via adjustment of laser duty cycle~\cite{Vasileska2020}.  Other applications of this instrument topology are found in~\cite{Fox2017, Lane2020}.

Hooper, using two on-axis cameras in a two color pyrometry setup, measure maximum temperatures, thermal gradients, and cooling rates, reported as $4000$~K, $20$~K/\textmu m, $40$~K/\textmu s, respectively, as typical of LPBF of Ti-6Al-4V~\cite{Hooper2018}.  Clearly, this substantiates the need for high dynamic range, fine spatial, and high temporal resolution to resolve fusion process dynamics.  In a related endeavor, Ma and colleagues investigate the meltpool dynamics of 316 stainless steel and report similarly extreme values~\cite{Ma2022}.  Finally, Vecchiato et al.~use imaging pyrometry to track solidification front velocity~\cite{Vecchiato2020}.  This is shown to be a function of laser parameters, and is presented as a mechanism for local control of component microstructure.

\subsubsection{Off-Axis Photodiode Techniques}

Off-axis application of a single pixel detector with a stationary field of view is reported by Bisht et al., who use a germanium photodiode sampled on a $20$~\textmu s period that observes radiance over the entire build area~\cite{Bisht2018}.  This work inversely relates the number of transients per volume of fused material to the ductility (elongation at break) of Ti-6Al-4V tensile test specimens.  Coeck and colleagues apply the same approach, albeit with two radiometers disposed on opposite sides of the laser delivery optics that are sampled at $50$~kHz~\cite{Coek2019}.  With processing to reject false positives, 92\% of pores are detected with an effective diameter of approximately $160$~\textmu m or greater.  An instrument by Dunbar and Nassar also features two photodiodes, though configured to measure a spectral line-to-continuum ratio~\cite{Dunbar2018}. This signature is also roughly correlated to density.

\subsubsection{On-Axis Photodiode Techniques}

To achieve some degree of spatially-resolved information, it is far more common that photodiode instruments are engineered to observe material fusion along the same path as the laser.  While not calibrated to temperature, a two-color photodiode pyrometer is described in~\cite{Doubenskaia2010}, which shows increasing heat accumulation as a series of six adjacent hatches are fused.  Follow-up work demonstrates that pyrometer signal levels additionally change with hatch spacing, hatch distance, and powder layer thickness~\cite{Pavlov2010}, indicating that process signatures are highly specific to the selected process parameters.  Work led by Okaro~\cite{Okaro2019} uses a machine learning approach on data from two photodiodes (filtered to $700-1050$~nm and $1100-1700$~nm); they are able to classify successful fabrication of tensile test specimens with 77\% accuracy, defined as an ultimate tensile strength of better than 1400~MPa.  In related work with the same instrument and analysis approach, test cubes are classified as greater or less than $99$\% dense with $93.5$\% accuracy against an optical microscopy ground truth~\cite{Jayasinghe2022}.  Closed-loop process control using on-axis photodiode data is the subject of~\cite{Renken2018} and~\cite{Renken2019}.  Results demonstrate mitigation of overheating at sharp reversals in laser direction (i.e., at the end of one hatch and beginning of the next).  A finite-element model is also used to deduce feedforward commands, which is shown to further augment controller performance.  Finally, using a significantly more complex instrument, work led by Lough uses a visible spectrometer that is analogously integrated into a LPBF machine~\cite{Lough2020}.  Selective emissions in the plume are shown to depend on laser power, cover gas, and pressure.

\subsubsection{Hybrid Instruments}

Hybrid instruments that combine at least two of the above approaches are a common way to combine the benefits of high spatial and temporal resolution.  One particularly enduring instrument of this sort, comprising a high-speed NIR CMOS camera and large-area silicon photodiode, by Kruth and colleagues~\cite{Kruth2007Feedback}.  Therein, rudimentary control of laser power is demonstrated to maintain constant radiance when transitioning from printing fully-supported material to printing overhanging features, though bandwidth is a noted limitation.  Craeghs and coworkers apply this instrument to a variety of tasks including: system identification in~\cite{Craeghs2010}; mitigation of scanning of acute corners and entrainment of extra material at the perimeter of a component in~\cite{Craeghs2011}; and a rigorous investigation of the thermal effects of support features in~\cite{Craeghs2012}.  An effort led by Clijsters demonstrates the ability to correlate large ($100$~\textmu m) porosites to a CT baseline~\cite{Clijsters2014}.  Further still, this instrument resurfaces as a commercial product in work led by Kolb in~\cite{Kolb2018} and~\cite{Kolb2019}. The former study attempts to compare both photodiode and camera signals to component porosity, yet concludes "the commercial melt pool system is not capable of detecting geometrical deviations or porosity precisely."  Nonetheless, the latter study is able to correlate surface roughness and balling defects using just the camera data stream.  Another instrument in this class is described by Thombansen, Gatej, and Pererira in~\cite{Thombansen2014Lens} and~\cite{Thombansen2014Tracking}, who address effects of chromatic aberration in the optical path.  Primarily, they demonstrate detection of overheating while scanning a powderless, grooved sample from which they estimate that detection of $400$~\textmu m-scale defects is plausible.  Chivel and Smurov use a combination of a high speed CCD camera and two-color, InGaAs photodiode pyrometer to study overheating when printing overhanging material, as well as Rayleigh-Tailor (balling) meltpool instabilities~\cite{Chivel2011,Chivel2013}.  Finally, a sensor fusion algorithm by Goosens and Van~Hooreweder creates a virtual sensor that is comprised of both photodiode and camera measurements, and validate the output of this virtual sensor against ex-situ measurement of meltpool depth~\cite{Goosens2021}.  Comparable instruments are also found in~\cite{Zhirnov2015, Montazeri2018, Forien2020, Thanki2022}.

\subsubsection{Limitations}

There are several frequently-identified impediments to reducing the minimum LPBF flaw size detectable with optical process monitoring techniques.  For example, on-axis imaging is negatively impacted by the innate characteristics of the f-theta lens.  By definition, f-theta lenses generate a high level of barrel distortion to linearize the $f \tan\theta $ characteristic of conventional (imaging) lenses~\cite{Smith2005}.  The linear angle-position mapping simplifies defining laser scan trajectories; however, images collected through an f-theta lens are distorted~\cite{Craeghs2011, Thombansen2014Lens}.  Moreover, optical strategies for achieving the requisite degree of distortion lead to poor focusing of wavelengths different from the design wavelength (i.e., they feature high chromatic aberration)~\cite{Smith2005, Thombansen2014Lens}.  Thus, the wavelengths used for monitoring must be close to that of the process laser, and not necessarily those optimal for process interrogation~\cite{Craeghs2012}.  Likewise, the off-axis approach necessarily induces prospective distortion, which is ideally corrected in post-processing~\cite{Le2021Monitoring} or via calibration~\cite{Lane2016}.  Finite depth of focus is additionally identified as a limitation to specimen size when using this approach~\cite{Grasso2016}. 

\subsection{Aperture Division Multiplexing}

\begin{figure*}[htbp]
\begin{center}
	{
	\includegraphics[trim = {0.95in 2.5in 1in 2.5in}, clip, keepaspectratio=true]{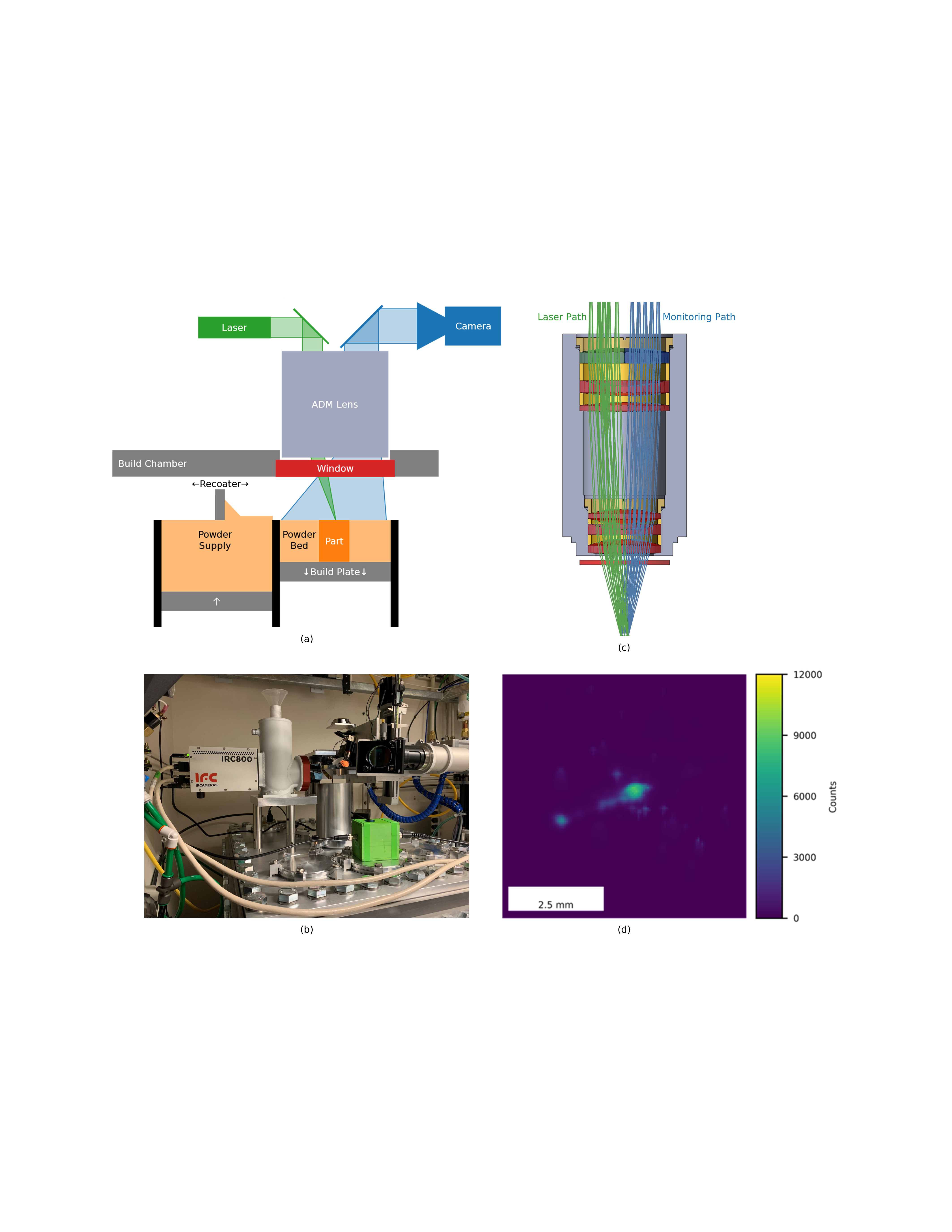}
	}
\end{center}
\vspace{-11pt}
\caption{Laser powder bed fusion using an ADM optic. (a) Schematic illustration of ADM, showing optical paths for both laser delivery and process monitoring through a common optic.  (b) ADM lens and camera attached to the top of the LPBF testbed, along with the laser boom (collimator, turning mirror, and galvanometers) entering from the right. (c) Overlay of optical ray trace and mechanical CAD for the ADM lens, showing lenses separated by brass spacers and secured in the lens tube with threaded spacers. (d) Typical meltpool image collected during printing.}
\label{fig:Fig2}
\end{figure*}

We present aperture division multiplexing (ADM) as a novel strategy for optical access to the LPBF process. We demonstrate the use of ADM for MWIR microscopy, enabling extraction of process signatures that are quantitatively correlated to pores in the size range known to nucleate fatigue cracks in LPBF components.  Figure~\ref{fig:Fig2}a schematically illustrates two optical paths through a common ADM lens: a first path is dedicated to directing and focusing the laser light to the build area of a LPBF machine and a second optical path uses a different portion of the lens to create an optical relay for process monitoring.  Combined with a high-speed MWIR camera, low distortion and high resolution ($50$~\textmu m) imaging is achieved, along with the high light throughput necessary for high temporal resolution.  Proving this instrument in a production-relevant context using a LPBF testbed, process signatures are extracted from video collected during fabrication of a metal test artifact.  These are compared to ex-situ micro-CT measurement of component density and detection probabilities are determined for pores $4.3$~\textmu m and larger.

\section{Methods}
\label{sec:Methods}

\subsection{LPBF Testbed}

This work utilizes a bespoke LPBF testbed designed to enable the ADM-based monitoring strategy.  Perhaps the most critical constraint is a limitation to the distance between the main plate and enclosure lid to no more than $60$~mm.  On one hand, this sets a lower bound on the working distance of any optics used in laser delivery or process monitoring.  On the other hand, this distance must be sufficient to accommodate a powder spreading (recoating) mechanism.  A second, related design requirement is that the top of the build enclosure need not be removed to operate the machine.  The optical equipment described herein attaches to mounting features on the lid and, as its installation requires a considerable effort in focusing and alignment steps, it is desirable to operate the testbed without removing them. This is met through the unconventional design explained in \S~\ref{sec:SupTestbed}, wherein build plate installation and part removal occur from the front of the machine. Otherwise, this instrument replicates the functionality of commercial equivalents, and has a build piston, piston-fed powder supply, and compliant-blade recoating mechanism, all actuated with industrial-automation-grade servo motors.

The top of the enclosure of the LPBF testbed is visible in Fig.~\ref{fig:Fig2}b, along with the laser scan head that is suspended on a boom above it.  This assembly delivers the laser light (the reader is referred to~\cite{Baker2017, Gibbs2018, Griggs2021, Kutschke2023} for additional design details) from a $500$~W, $1.075$~\textmu m fiber laser (redPOWER qube, SPI).  The laser is a single-mode fiber laser with a beam divergence of $82$~mrad and, in conjunction with a 250 mm collimator (Coherent PN 106402X01), produces a beam with a diameter of $d=2\times NA\times f$ or $19.5$~mm.  The figure also shows that light from the collimator is redirected by a turning mirror before a galvanometer set (Thorlabs PN QS20XY) that performs laser scanning.

All machine functions are automated with an NI-cRIO-9039.  A layer cycle begins with a powder spreading operation and, once complete, a gas knife is enabled two seconds prior to beginning layer fusion. Next, the cRIO synchronously operates the laser and galvanometer mirrors of a commands on a $10$~\textmu s timebase.  Once fusion of a layer concludes, the gas knife is operated for an additional two seconds before the next recoating operation begins.  This control signal is also used to trigger the camera for the same time period.

\subsection{Aperture Division Multiplexing Lens}

\subsubsection{Optical Design}
\label{sec:MethodsADMOpticalDesign}

The optical path dedicated to laser delivery requires a clear aperture of $20$~mm diameter and a laser damage threshold greater than $500$~W distributed over the clear aperture, while also bringing $1.07$~\textmu m light to an $\approx 70$~\textmu m diameter spot.  The imaging path is designed to operate with a previously obtained MWIR camera equipment that is described below.  This requires a $22$~mm diameter clear aperture and sets the focal length of the ADM lens as $125$~mm to achieve a geometric resolution of $50$~\textmu m.  Design wavelengths for this path are nominally $1.2-2.4+$~\textmu m, yet optical materials are chosen to allow monitoring at longer wavelengths with minimal redesign.  Performance metrics for both paths are specified on a $6$~mm diameter field-of-view\footnote{This, in effect, results in lower-yet-sufficient performance over a $\approx 10$~mm diameter field-of-view.}.  Finally, an optical window is required to protect the ADM lens from the build environment and to maintain the atmosphere in the LPBF testbed.  It is specified as a $4$~mm thick, $76.2$~mm diameter piece of calcium fluoride (ultimately surfaces~13 and~14 in Table~\ref{table:BroadbandPrescription}), with the bottom-most surface placed $60$~mm above the build plate (co-planar with the bottom surface of the LPBF testbed enclosure lid).  This bounds the working distance of the ADM lens.  A final requirement is the ability to independently adjust the laser focus size without affecting the imaging performance.  For this purpose, the first optical element of the ADM lens design is split such that the radii of the first surface in each optical path can be independently changed.

A Petzval-inspired design, with each optical path decentered by $17.5$~mm, is used as a starting point to meet these objectives.  Traditionally, these comprise two pairs of achromatic doublets, where each bends the axial ray by the same amount (i.e., optical power is split roughly evenly)~\cite{Smith2005}.  In our design, positive focusing power comes from the calcium fluoride elements and chromatic aberration is controlled with the negative fused silica elements (see additional notes on optical materials and coatings in \S~\ref{sec:SupADM}).  In a two-doublet design, however, achieving the required optical power with calcium fluoride requires strongly-curved surfaces due to the low refractive index of this material.  As such, the positive elements are split and placed on opposite sides of the negative elements in the present design, making two three-element groups.  Distributing the positive optical power across four elements instead of two implies lower radii of curvature and, hence, lower aberration.  

The other half of the imaging relay comprises a high-speed MWIR camera (IRCameras IRC806HS) equipped with a $50\,\mathrm{mm}$, f/2.3 lens (Stingray Optics PN SR1936-A01).  The camera comprises a $640 \times 512$~px InSb (indium antimonide) detector, sensitive from $1$ to $5$~\textmu m.  Pixels are $20$~\textmu m square and have a well depth of approximately $7$~million photoelectrons.  An f/2.3 cold stop is installed within the vacuum dewar, though a cold filter is not. 

A merit function is used to evaluate and computationally optimize the ADM lens design in Zemax software, starting from this initial form, where the laser path and imaging path are simultaneously considered using two corresponding model configurations.  Performance of the laser path is evaluated directly, where light from the laser is collimated in the simulated object plane and comes to a focus in the image plane.  Evaluating the imaging path at the same time is more complex because it is most-straightforward to consider light propagating in the same direction in both model configurations (i.e., in the reverse direction as compared to how this path is used in the physical optic).  Therefore, a pupil-matching strategy is used to define the rays traced in this model configuration in this reverse direction, in view of the nominal performance metrics of the Stingray lens.  Geometric considerations are also included to ensure the manufacturability of the elements comprising the ADM lens.  In general, performance metrics of the imaging path and satisfying geometric constraints are weighted more heavily, as the other LPBF process parameters (namely laser power and scan speed) can be tailored to compensate for considerable variation in the as-built laser spot size.  A summary of merit function objectives is given in~\S~\ref{sec:SupADM}.  After optimization, a tolerance analysis is performed (also described in~\S~\ref{sec:SupADM}) for predicting as-built performance.  This is met, in part, using a lens tube strategy (depicted in Fig.~\ref{fig:Fig2}c) to constrain the optical elements that make up the ADM lens. 

\subsubsection{Camera Fixture}

Figure~\ref{fig:Fig2}b also shows the relative position where the camera is fixured with respect to the ADM lens. Thorlabs optical posts (PNs RS2 and RS3 to reach $5$~in. total length) are used to interface mounting features on the LPBF testbed environmental enclosure to a fixturing plate.  Because the MWIR camera features a pour-filled liquid nitrogen dewar, it is necessary to fold the optical path such that the camera remains upright while effectively looking downwards.  A gold-coated turning prism is used for this purpose (Edmond Optics PN 47-031).  To provide mounting features, the prism is adhesively bonded to a two-part assembly that interfaces it with a 2 axis goniometer (Thorlabs PN GN2).  The stationary side of the goniometer interfaces with the same plate used to fixture the camera by a second machined component and Thorlabs optical posts (PN RS4).  Accordingly, adjustment of the goiniometer rotates the prism with respect to the camera and thereby moves its field of view relative to the center of the build area.

\subsection{Camera Parameters}

The camera is interfaced with a PC separate from the rest of the printer infrastructure via twin Camera-Link cables and a NI PCIE-1433 frame grabber.  It is configured to observe a trigger signal in its operational software (IRCameras WinIRC), as previously mentioned, to avoid collecting meaningless data during the recoating cycle.  This is accomplished by splitting the gas knife enable signal, connecting it to an optoisolator (SparkFun PN BOB-09118) that provides signal level shifting and protection, and finally to the BNC trigger IO port of the camera.  To adjust the remaining imaging parameters before starting a print, the machine code for the first layer of the part is manually scanned on the bare build plate.  From these data, the exposure time is set as a balance between avoiding saturation under typical conditions and making optimal use of the dynamic range of the detector.  An extremely short exposure time of $0.7$~\textmu s is made possible using the ADM lens, and is useful here to prevent rapid changes in the scene from causing a burring effect.  Windowing is also applied, where only a subset of the pixels are read out.  This is desirable for two related reasons in that it lowers the amount of data that needs to be transferred and stored, and also allows the frame rate to be increased.  As set here, capturing the entire cross section of the present print allows a frame rate of $1250$~Hz.

\begin{figure*}[htbp]
\begin{center}
	{
	\includegraphics[trim = {0.8in 2.75in 0.95in 2.75in}, clip, scale=1, keepaspectratio=true]{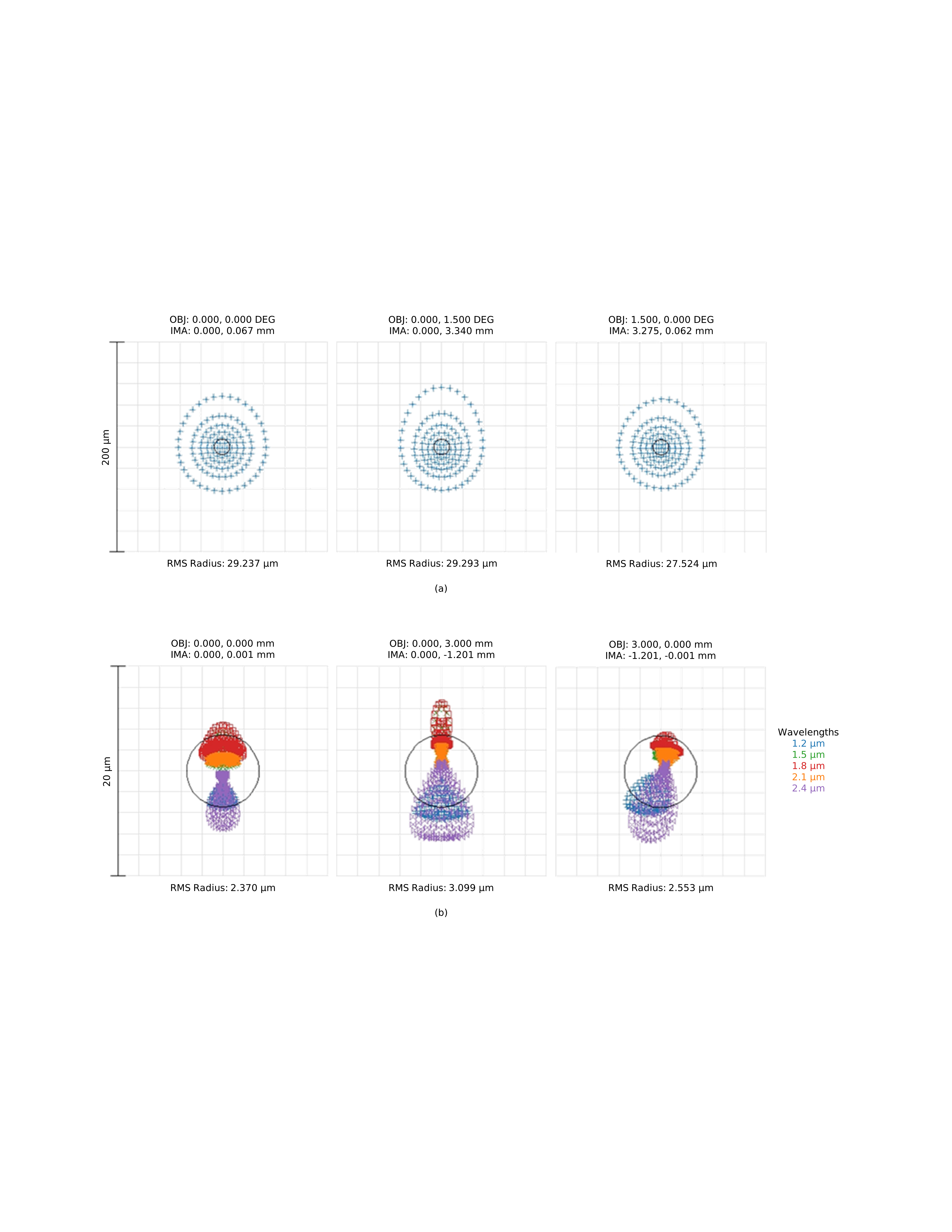}
	}
\end{center}
\caption{Spot diagrams for the ADM lens design.  (a) Spot diagram for three representative fields for laser light delivered to the build plane.  (b) Spot diagram for three representative fields for light from the build plane imaged onto the camera sensor.}
\label{fig:Fig3}
\end{figure*}

\subsection{Build Planning and Printing}

Autodesk Netfabb software is used for print planning.  A part is first sliced on a $30$~\textmu m interval and the perimeter of each slice is offest in the inward direction by $35$~\textmu m, or roughly half the laser spot size.  These slices are first used to bound the start and end points of the infill hatches, each separated by $30$~\textmu m, and are all scanned in the same direction.  The hatch direction is rotated by $67^\circ$ every layer.  Finally, a copy of the offset perimeter is placed at the end of the hatch command list for scanning the perimeter of the part, such that these commands can be exported as a .cli in that order.  Machine code generation follows, using a .cli interpreter Python script.  The final scan parameters are set here, including a scan speed of $250$~mm/s and laser power of $100$~W that are both used for the infill hatches and perimeter (polyline) commands.  A $0.175$~radian ($\approx 10^\circ$) rotation is also applied, such that the recoater does not encounter the entire leading edge of the component at the same instant.

Final print parameters arranged at runtime.  316~stainless steel powder (Carpenter Technology, $15-45$~\textmu m) is used as the feedstock.  To prevent print failure from a powder short-feed, the powder piston set to express enough powder for a $50$~\textmu m layer despite the $30$~\textmu m layer height specified.  Ultra-high purity argon (Airgas, 99.999\%) is used as the cover gas.

\subsection{Density Characterization}

Micro-CT (computed tomography) is used to provide ground truth of component density. Micro-CT is performed using a Zeiss Xradia 620 Versa CT machine. Source settings include a tube potential of $160$~keV, power of $25$~W, and to apply a filter (HE6).   The imaging procedure collects $1601$~projections, using the $0.4\times$ detector set to a $6$~s exposure time.  In sum, the resulting reconstruction is provided at $4.3$~\textmu m resolution (voxel size).  One challenge to reconstructing the density of a specimen via computed tomography is beam hardening; unresolved, this complicates the following dataset alignment and pore identification tasks.  A slightly modified version of Otsu's method~\cite{Otsu1979} is applied to remove this reconstruction defect from the dataset by binary thresholding.

\section{Results}

\subsection{ADM Lens}

\begin{figure*}[htbp]
\begin{center}
	{
	\includegraphics[trim = {0.8in 5.55in 0.8in 2.2in}, clip, scale=1, keepaspectratio=true]{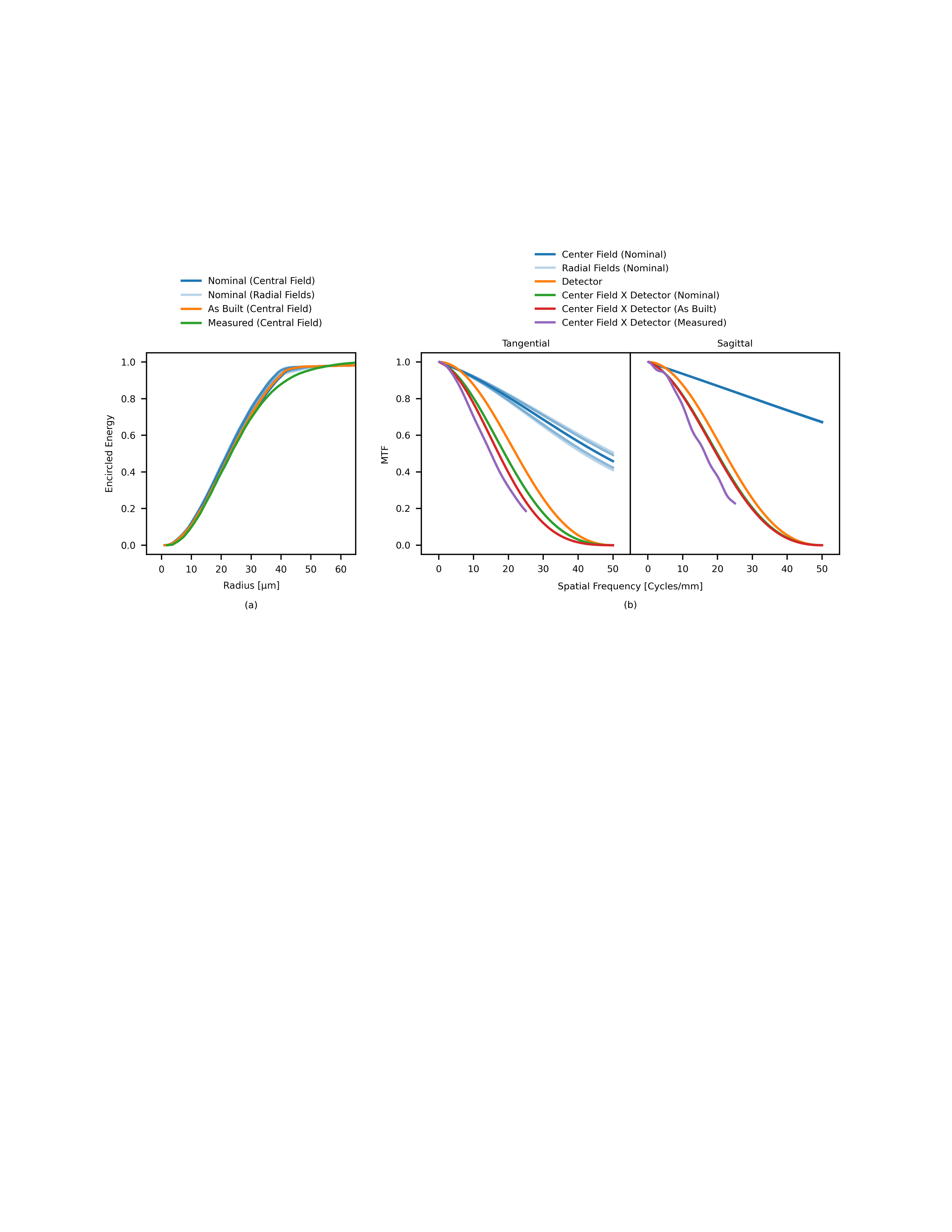}
	}
\end{center}
\caption{Comparison of simulated and measured ADM lens performance.  (a) Plot of encircled energy, showing that the as-measured D86 (laser spot size) is $77$~\textmu m and corresponds closely to the expected profile.  (b) Modulus of the optical transfer function for the simulated ADM lens, detector, and composite system.}
\label{fig:Fig4}
\end{figure*}

The goal of ADM lens design for LPBF is to engineer two optical paths through a single lens: one path that focuses the laser light and one path that enables high-resolution process monitoring.  Following the method described in \S~\ref{sec:MethodsADMOpticalDesign}, Zemax is used to simultaneously evaluate the performance of both optical paths for design optimization.  The resulting ADM lens prescription is given in Table~\ref{table:BroadbandPrescription} and performance metrics are summarized in Figs.~\ref{fig:Fig3} and~\ref{fig:Fig4}.  Laser spots for representative fields are presented in Fig.~\ref{fig:Fig3}a, showing uniform performance over the specified field of view. The data are alternatively presented as encircled energy in Fig.~\ref{fig:Fig4}a, from which it can be seen that a D86 of $76$~\textmu m is expected.  Final imaging performance is evaluated with a second optical model, where the model described in \S~\ref{sec:MethodsADMOpticalDesign} is reversed, the Stingray lens is added as a paraxal element of nominally identical parameters, and rays are traced from the build (object) plane to the detector (image plane).  Selected spots resulting from this process are shown in Fig.~\ref{fig:Fig3}b, where the RMS spot radii are expected to be an order of magnitude smaller than the pixel dimensions and the design is therefore expected to achieve high-quality imaging.  While not used as an optimization metric, the resolution of the complete optical relay is alternatively quantified using the modulus of the optical transfer function (MTF) in Fig.~\ref{fig:Fig3}b.  Practically, these curves consider contrast as a function of spatial frequency or feature size and provide a more nuanced understanding of system resolution than a simple computation of geometric magnification.  To explain in a different way, MTFs for the as-built system are deduced using the slanted-edge test, in which an image of a razor blade is recorded and system performance deduced from how severely the edge is blurred. 

Figure~\ref{fig:Fig3}b makes a number of comparisons using MTFs.  Blue lines in the plots are generated in Zemax using the nominal optical path, including the ADM lens and Stingray lens (again modeled as a paraxial element).  The orange curve is the MTF of the detector, which, due to its finite pixel size, cannot resolve infinitely small features.  It is simply computed as $\mathrm{MTF_{DET}} = |sinc(\xi w)|$, where $\xi$ is spatial frequency and $w$ is pixel pitch.  Much of the utility of these curves is that the MTF of a composite system can be predicted by multiplying the MTFs of the individual parts.  As such, the green curves predict the performance of the combination of the modeled optical path and detector.  

\begin{table}[bt]
\centering
\footnotesize
\caption{ADM lens prescription.  Surface 1A defines the D shaped lens used for optical monitoring, and Surface 1B defines the equivalent surface for laser delivery.}
\label{table:BroadbandPrescription}
\begin{tabular}{@{}*5c@{}}
  
  \toprule[1.5pt]
  \multicolumn{1}{c}{\head{Surface}} &
  \multicolumn{1}{c}{\head{Radius}} &
  \multicolumn{1}{c}{\head{Thickness}}&
  \multicolumn{1}{c}{\head{Material}}&
  \multicolumn{1}{c}{\head{Semi-Diameter}}\\
  
  \cmidrule{1-5}
 
    1A & 190.794 & 13.034 & CaF\textsubscript{2} & 35.000\\
    1B & 193.933 & 13.034 & CaF\textsubscript{2} & 35.000\\
    2 & Infinity & 19.420 & - & 35.000\\
    3 & -132.588 & 8.244 & Fused Silica & 35.000\\
    4 & -258.974 & 5.525 & - & 35.000\\
    5 & 282.615 & 7.858 & CaF\textsubscript{2} & 35.000\\
    6 & -1097.974 & 82.634 & - & 35.000\\
    7 & 123.936 & 12.645 & CaF\textsubscript{2} & 28.000\\
    8 & -102.040 & 4.725 & - & 28.000\\
    9 & -86.569 & 4.501 & Fused Silica & 28.000\\
    10 & -2352.203 & 3.517 & - & 28.000\\
    11 & 92.640 & 12.645 & CaF\textsubscript{2} & 28.000\\
    12 & -403.854 & 5.000 & - & 28.000\\
    13 & Infinity & 4.000 & CaF\textsubscript{2} & 38.100\\
    14 & Infinity & 60.000 & - & 38.100\\

  \bottomrule[1.5pt]
\end{tabular}
\end{table}

\begin{figure*}[htbp]
\begin{center}
	{
	\includegraphics[trim = {1.2in 2.1in 1.2in 2.1in}, clip, scale=1, keepaspectratio=true]{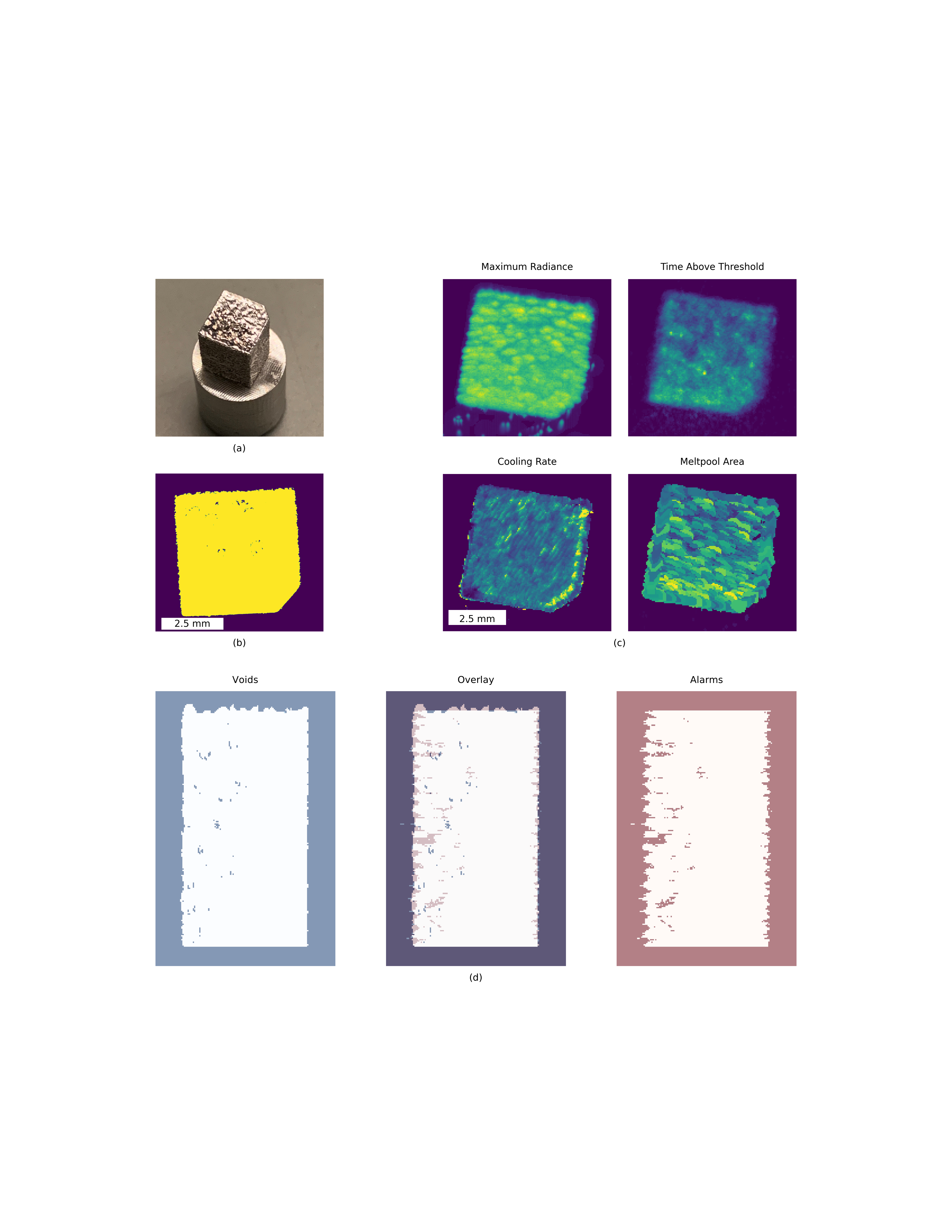}
	}
\end{center}
\vspace{-11pt}
\caption{Products resulting from the fabrication of the test artifact.  (a) As-built image of the test artifact prepared for micro-CT.  (b) Representative slice of the micro-CT dataset.  (c) Process signatures extracted from the video data at approximately the same location in the test artifact as (b). Color scales are qualitatively adjusted to clarify variation across the component cross section. (d) Thresholded density and alarm datasets, along with overlay (center), showing that voids and corresponding alarms are not necessarily co-incident.}
\label{fig:Fig5}
\end{figure*} 

\subsubsection{Qualification}
\label{sec:MethodsQualification}

Due to the deliberate change in optical material described in \S~\ref{sec:SupADM}, the physical ADM lens is not qualified against the nominal models.  Rather, a second pair of optical models use the as-built dimensions and materials (i.e., inspection report data and actual dispersion data for Corning~7980 fused silica~\cite{Corning2015}).  Each path is evaluated independently and therefore qualification of the as-built assembly proceeded in two stages.

% paragraph{Laser Path}

Performance of the laser path is quantified directly on the LPBF testbed.  For this purpose, a board level camera (The Imaging Source PN DMM 37UX226-ML with a Sony IMX226 sensor) is fixtured within the printer such that the sensor lies in the nominal build plane.  To reduce the laser power to a suitable level for the camera sensor, the turning mirror in the laser head is replaced with a beam sampler (Thorlabs PN BSF20-C) and two neutral density (ND) filters (Thorlabs PNs  NDUV2R40A and NENIR40A) are additionally inserted up stream of the galvanometer mirrors.  Leveraging the $1.85$~\textmu m pixel pitch of this sensor, the laser spot size can then be imaged directly.  Figure~\ref{fig:Fig4}a shows the result of integrating a typical as-measured laser spot as a function of distance from its centroid, arriving at a plot of encircled energy comparable to those generated from the optical design software.  The measured D86 is $77$~\textmu m, closely matching the expected performance.

% \paragraph{Imaging Path}

As mentioned above, resolution of the optical path is quantified by measurement of the system MTF using the slanted-edge method.  Returning to Fig.~\ref{fig:Fig4}b, the red curve is the MTF of the as-built Zemax model and performs slightly worse than the nominal design as a result.  Purple curves give the results of the experimental system MTF measurements and are most fairly compared to the red curves.  These lag slightly below the theoretical performance for two reasons; namely, mechanical imperfection due to the machined components in the ADM lens assembly are not considered and performance of the Stingray lens is over-approximated.  While the contrast needed to resolve a specific feature or phenomenon is somewhat situation dependent, contrast is roughly 20\% at the Nyquist limit (or the highest spatial frequency directly resolvable with a detector of a given pixel size in view of system magnification) and it is therefore expected that features at the geometric resolution of $50$~\textmu m are resolvable.

\subsection{Test Artifact}

To demonstrate the performance of the ADM along with MWIR microscopy, a simple cubic test artifact ($5 \times 5 \times 6$~mm) is printed, shown in Fig.~\ref{fig:Fig5}a.  Its geometry is selected both to fit within the best part of the ADM lens field-of-view and to enable a small voxel size when performing micro-CT.  The $6$~mm dimension is the height of the component, and provides a $1$~mm margin to allow the process to reach steady-state though the first several layers.  The figure also shows that a $1\times 1$~mm chamfer is added along one vertical edge of the part.  This feature serves as a fiducial marker that is visible in both the camera and micro-CT data; accordingly, the datasets can be unambiguously co-registered.  Figure~\ref{fig:Fig2}d shows a typical frame (image) recorded during its manufacture, using the imaging approach described above.  The large spot at the center is coincident with the laser spot, and the rest of the meltpool is visible as a short tail to the lower left.  One additional bright spot is visible farther in this direction, which is a location of overheating and perhaps a balling defect.  Other bright regions in the image are caused by spatter.  

\subsection{Process Signatures}

Figure~\ref{fig:Fig5}c shows four process signatures that are extracted from the camera dataset for each layer of the part: time above threshold, maximum radiance, meltpool area, and cooling rate.  Time above threshold is the simplest computationally, where the number of radiance values above $4000$~counts along the temporal axis are tallied for each pixel.  The $4000$~count value was  chosen from a casual inspection of the video data, like many other parameters in these routines.  

The remaining signatures rely on estimating the location of the meltpool via center-of-mass calculation.  Maximum radiance for each pixel is found by first determining the subset of radiance measurements where the laser spot is within a $10$~pixel radius of the pixel under consideration, then selecting the maximum value of this subset.  This step helps to reject clutter from hot spatter particles.  Meltpool area (at last estimated melting event) uses the same clutter-rejection strategy to find the last time that a pixel exceeded $5000$~counts.  At that time, the number of adjacent pixels that are also above this threshold are tallied.  

A proxy for cooling rate is the most complex metric extracted from these data and begins with finding radiance maxima that satisfy three criteria.  First, the maxima must occur when the meltpool center is within $10$~pixels of the pixel being assessed.  Second, the peak must be at least $6000$~counts.  Third, the following two points after the peak must be at least $1200$~counts to ensure genuine melting events are selected.  Melting events are truncated at $50$ frames, except in cases where a subsequent melting event occurs in that time and they are truncated to when the second event begins in this case.  Next, the melting events are time-aligned and averaged to arrive at an average cooling curve.  An exponential equation of the form $a\cdot e^{-bt}$ is fit to the average cooling curve, where $b$ is the cooling rate.

\begin{figure*}[htbp]
\begin{center}
	{
	\includegraphics[trim = {1in 1.11in 1in 1.06in}, clip, scale=1, keepaspectratio=true]{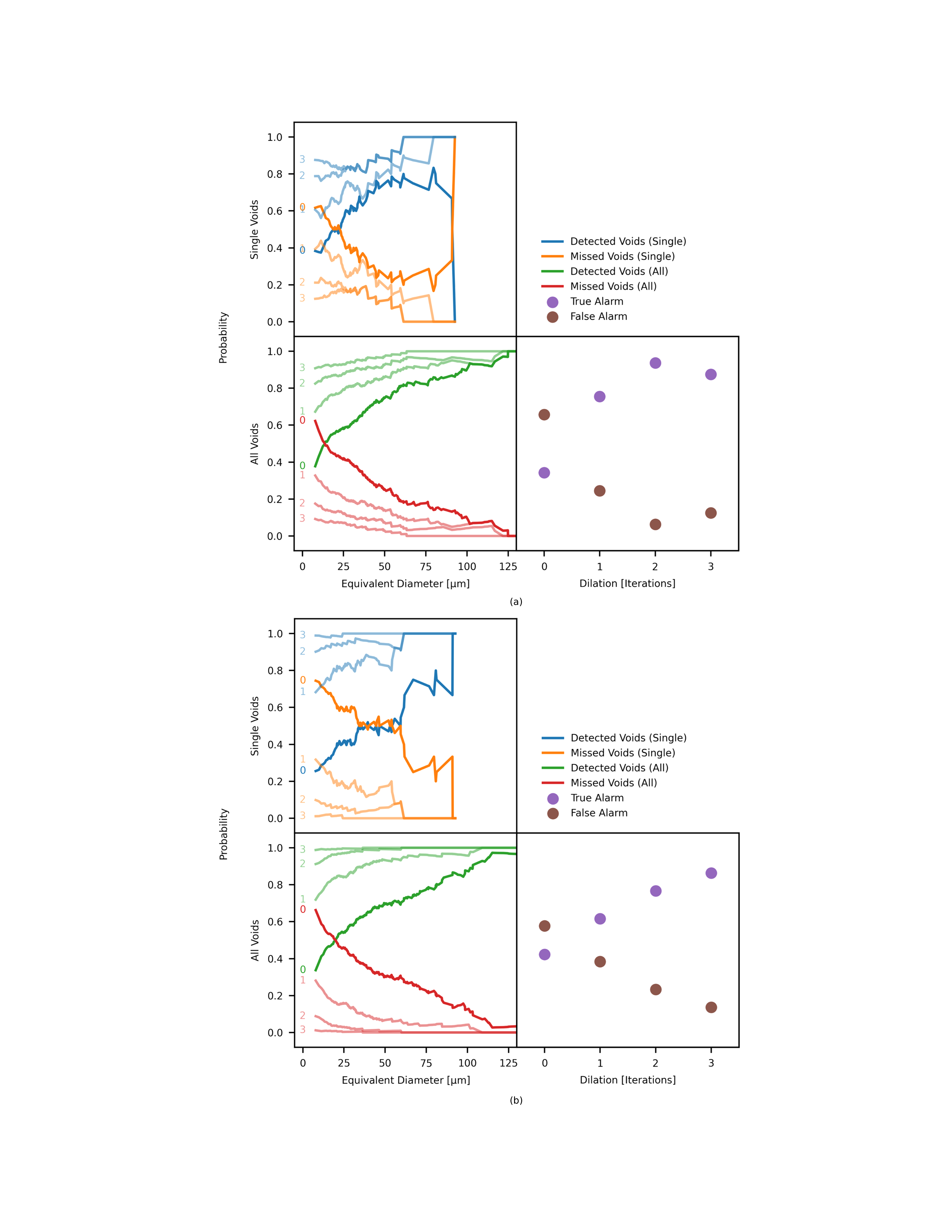}
	}
\end{center}
\vspace{-11pt}
\caption{Detection probabilities using time above threshold as a process signature.  (a) A low threshold ($\leq 8$~ms) is applied to indicate alarms.  In the left-hand panels, the solid lines represent thresholds without dilating the detections and the faint lines are labeled with the number of dilations applied with detection probability improving as a result.  (b) Identical to (a), except that a high threshold ($\geq 64$~ms) has been applied.}
\label{fig:Fig6}
\end{figure*}

\subsection{Dataset Alignment}

A dataspace transformation is used to project micro-CT voxels into the process signature dataspace; mathematically, this mapping is performed with a homogeneous transformation matrix (HTM) that performs three functions.  First, the HTM provides for three rotations, as the test artifact is not rotated in precisely the same orientation in the micro-CT data as it is seen by the camera during printing.  Second, it provides for three shifts, as the test artifact is not perfectly centered in the micro-CT reconstruction, nor in the field of view of the camera.  Third, it accommodates the difference in scale, where micro-CT voxels are $4.3 \times 4.3 \times 4.3$~\textmu m$^3$ and process signature voxels are taken to be $50 \times 50 \times 30$~\textmu m$^3$ (the product of camera resolution and layer thickness).  

Optimizing the HTM from a rough starting point faces two complicating factors.  First, the functional relationship between process signature values and component density values is unknown and reasonably expected to be nonlinear.  Alignment methods that assume a linear relationship between such quantities, such as cross-correlation, generally perform poorly in this circumstance~\cite{Collignon1995}.  Second, Figs.~\ref{fig:Fig5}b and~\ref{fig:Fig5}c show that the geometric features are, for lack of a better term, blobby. This makes conventional image alignment techniques that begin with feature extraction algorithms (e.g. Harris corner detector~\cite{Harris1988} or SIFT~\cite{Lowe1999}), ill-suited.  In contrast, intensity-based registration methods, while typically applied in niche applications, are an excellent fit to the task at hand.  Intensity-based methods function by optimizing a statistical similarity measure, frequently mutual information (MI) as is used here, across the values in all possible pairs of pixels (voxels)~\cite{Collignon1995, Viola1995, Pluim2003}.  Optimization of the HTM components using MI as an objective function is a delicate process; the Scipy implementation of Powell's method is used (see~\cite{Scipy2020} and~\cite{Powell1964}.).  Derivativeless optimization routines like this one are typically preferred to their quasi-Newton counterparts, as to avoid problems arising from ill-natured derivatives of the mutual information objective function~\cite{Pluim2003}.  Finally, it should be noted that the micro-CT dataset can be aligned to any of the process signatures described in the previous section; however, a single HTM determined using this approach and a sufficiently predictive process signature (e.g., maximum radiance) can generally be used with any other process signature without significant refinement. 

\begin{figure*}[htbp]
\begin{center}
	{
	\includegraphics[trim = {1in 3.25in 1in 3.5in}, clip, scale=1, keepaspectratio=true]{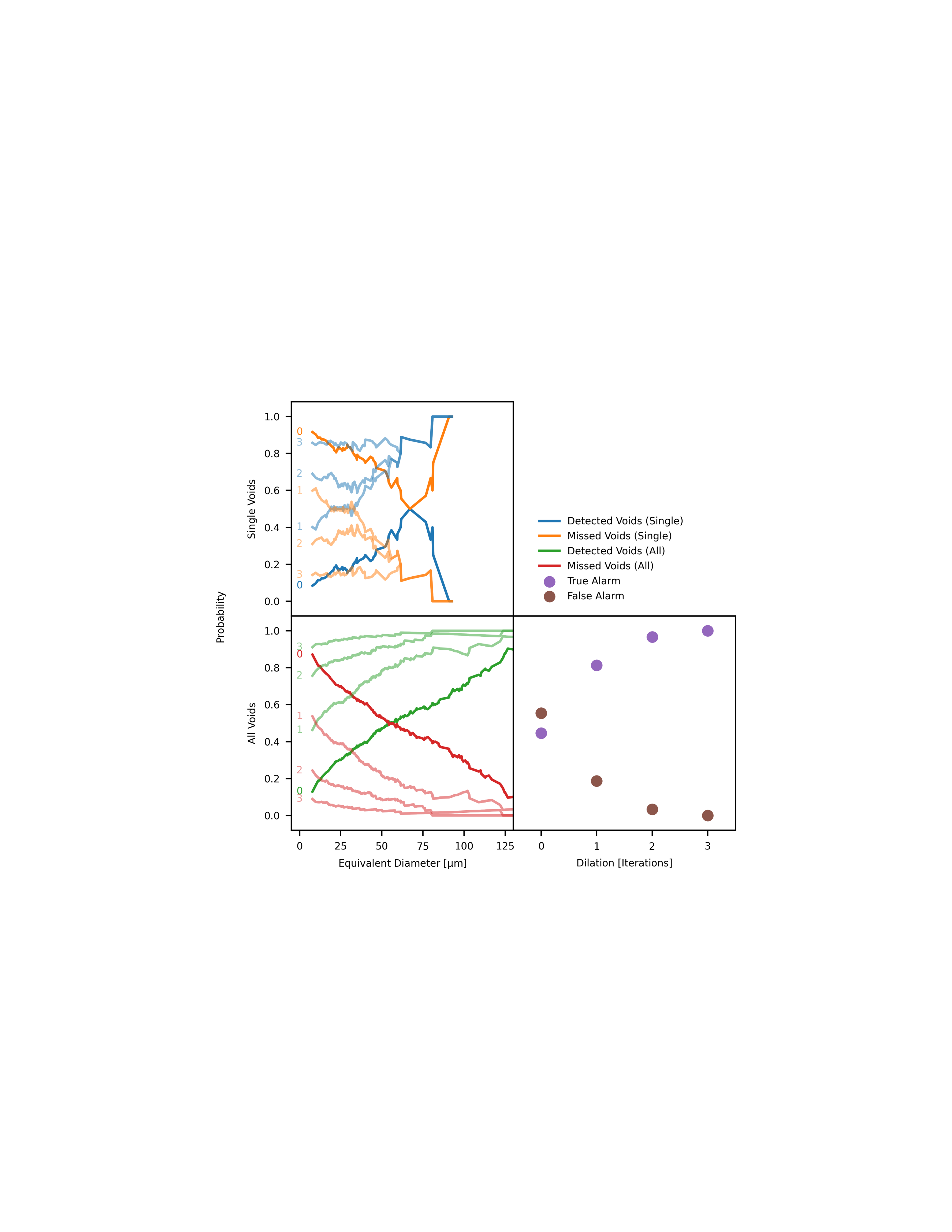}
	}
\end{center}
\caption{Detection probabilities using rate as a low-threshold ($\leq 0.1 \, \mathrm{ms}^{-1}$) process signature.  Definition of each panel directly correspond to those in Fig.~\ref{fig:Fig6}a.}
\label{fig:Fig7}
\end{figure*}

\subsection{Pore Detection Statistics}

A pore-by-pore and alarm-by-alarm analysis is conducted on the (aligned) density and signature datasets to evaluate the probability that pores of a certain size are detected or missed.  While perhaps self-explanatory, a pore is taken to be the maximal collection of adjacent micro-CT voxels judged to be void of material after thresholding.  There is a probability that such a pore is detected or missed.  Likewise, an alarm is the maximal collection of adjacent voxels in the process signature matrix that lie outside of a threshold range taken to indicate stable material fusion.  There is a probability that an alarm is true, meaning it positively identifies (overlaps with) a pore, or is false and indicates there is a flaw at a location where the material is actually fully dense.

Two other features are used to refine this analysis.  First, it is quite typical that an alarm does not occur in the same location as the pore.  For example, a pore may be indicated as a hot spot in the next layer, as is particularly evident in the overlaid dataset slices in Fig.~\ref{fig:Fig5}d.  This is studied by dilating the thresholded process signature array.  One dilation iteration makes the $26$ voxels surrounding an original alarm voxel also alarm voxels if they are not already so, expanding the extent of the alarm $1$~voxel in all directions.  Second, after projection, it is possible to identify a subset of voids that do not share any process signature voxels with another void.  To clarify, in the case of a plurality of voids within one process signature voxel, it is impossible to know which void(s) generated the alarm.  This subset of isolated voids, then, arguably provides a more accurate sense of void detection probabilities.  

We first consider the performance of time above threshold as a process signature to predict the presence of pores as identified by micro-CT.   Figure~\ref{fig:Fig6}a summarizes pore detection probabilities when generating alarms in locations of low time above threshold (i.e., less than $8\,\mathrm{ms}$).  It is hypothesized that a low time above threshold indicates porosity, either through application of insufficient energy to fully melt the material of from rapid dewetting of molten material during a balling event that exposes the underlying (cooler) material.  In the bottom left panel, the cumulative detection probability for voids of a given size and larger is plotted as a function of the equivalent diameter of the void.  The solid lines show that the probability of detecting all of the voids resolved in the micro-CT data is around $36$\% and climbs as only increasingly large pores are considered.  Allowing for a slight difference in the position of the void and signature by dilating the signature twice greatly improves performance: more than $80$\% of all voids are detected.  To provide cleaner insight, the upper left panel plots the same curves considering only single voids (i.e., that overlap with another void in any signature-space voxel).  These curves are rougher, particularly at large equivalent diameters where few voids contribute to the statistic.  Nonetheless, a similar pattern is observed, where the process signature performs poorly when expecting a 1-to-1 spatial correlation, but with two dilations again shows the ability to detect roughly $77$\% of the voids resolved in the micro-CT ground truth.  The final panel considers the probability that an alarm is true (indicating that it overlaps with a void) or is false (indicating a void where the component is fully dense).  Clearly, a low number of false alarms is desirable to avoid needlessly failing parts undergoing qualification and this is not the case if the signature is used directly (i.e., with no dilation).  With two dilations, the probability an alarm is false is $6$\%.  Interestingly, the probability of a false alarm goes up if a third dilation is applied for this signature.  This is rationalized by noting that positive alarms tend to be near each other, and therefore become fewer in number as they coalesce, whereas the false alarms are farther apart on average.

Time above threshold is also one of the only signatures that, in the present dataset, has significant predictive power when also used at the opposite extreme.  While not quite as effective as the prior case, Fig.~\ref{fig:Fig6}b shows that high time above threshold is also predictive of porosity and is characteristic of the balling-type defects frequently observed in the test artifact.  Specifically, once this defect has occurred, the large ball is slow to cool.  After three dilations, 98.9\% of pores are detected with a false-alarm rate of $14$\%.  In comparing this signature to the prior one, it should be noted that the probability of detecting pores can be traded for a lower rate of false alarms by changing the thereshold applied to the process signature.  For example, a time threshold greater than the $\geq 64$~ms used here can make the rate of false detections equal to those seen in Fig.~\ref{fig:Fig6}a.  However, fine pores are also made less likely to be detected in this case.

Figure~\ref{fig:Fig7} presents the equivalent analysis using areas of low cooling rate to indicate porosity.  Deviations in this process signature are hypothesized to arise from one of two mechanisms. Like high time above threshold, balling defects can be slow to cool due to the concentrated volume of hot material.  Alternatively, the rate of thermal conduction from the meltpool is reduced by pores proximal thereto, and this can more subtly reduce the observed cooling rate.  In either case, performance of this process signature is comparable to low time above threshold, where after two dilations roughly 70\% of isolated pores are detected with a false alarm rate of 3\% and these metrics improve with a third dilation.  In contrast to the time above threshold signature, it is interesting to note how rapidly the performance of low cooling rate increases as a function of the number of dilations, particularly in resolving pores smaller than approximately $40$~\textmu m in effective diameter.  This may indicate that this signature is comparatively sensitive to fine pores lying under the meltpool.

Similar plots for maximum radiance and meltpool size at last melting event are given in \S~\ref{sec:DetectionThresholds}.  Maximum radiance gives similar results to time above threshold, and this agrees with the notion that large amounts of overheated material cause balling defects that cool slowly. Meltpool size at last melting event is easily the worst-performing signature, which is not surprising in view of its coarse resolution.  However, it is most predictive of porosity when very low, perhaps indicating process instability about a pre-existing balling defect.

While these process signatures are compared here, two limitations to these findings should be noted.  First, inspection of the micro-CT data show that balling is the predominant source of porosity in the present study, and the efficacy of these signatures should be evaluated for keyholing and more typical lack-of-fusion porosity in future research.  Second, the LPBF process parameters chosen are very near the edge of the process window, as evidenced by the frequency and nature of porosity present in the test artifact.  Close examination of Fig.~\ref{fig:Fig5}d shows that pores predominately appear in the left half of the test artifact, corresponding to the leading edge of the component with respect to the powder spreading direction.  Density in this region is about $98.7$\%, lying below expectations for well-executed LPBF (it should be noted that the density of the entire component is considerably higher at $99.4$\% which is more typical of this process).  We hypothesize that LPBF with parameters better chosen for this specific geometry will generate process signatures with lower baseline variability.  Thus, deviations in the process signatures associated with porosity are expected to be made more obvious in a higher density component.

\section{Discussion}

Aperture division multiplexing (ADM) was fully demonstrated through the design and validation of a lens for simultaneous laser delivery and in-situ microscopy in LPBF, achieving high spatial resolution and high light collection efficiency. Comparing process signatures extracted from the camera data to ex-situ micro-CT measurement of test artifact density showed predictive power for voids as small as $4.3$~\textmu m diameter.  This characteristic dimension is small as compared to the minimum pore sizes that are capable of nucleating a fatigue crack, proving the future viability of process monitoring via ADM for certification of LPBF component fatigue life.

Future research may directly use these developments to achieve higher spatial and temporal resolution.  Two routes exist to decreasing the minimum resolvable feature size with the present ADM optic, namely using a camera detector with finer pixel size and increasing the focal length of the lens affixed to the camera (i.e., replacing the $50$~mm Stingray lens).  The MTF analysis of Fig.~\ref{fig:Fig4}b suggests that resolving features on the order of $10$~\textmu m is possible with such a change.  Temporal resolution may also be increased, most directly by subframing (trading field of view for increased frame rate at the same net data rate).  The $0.7$~\textmu s exposure time used here places a theoretical upper limit of $\approx 1.4$~MHz on frame rate, which all but ensures camera data bandwidth is the practical bound.

The success of this reference design also justifies a future iteration to improve ADM lens performance, where one of the greatest impediments to the present design is the nature of the chromatic aberration (or the lack thereof) that must be engineered into the lens.  Relieving this design pressure via one of the following options makes more optical surfaces available for correcting achromatic aberrations, critical to expanding the field of view, at the same level of system complexity.  One option is a change of optical materials.  At the present wavelengths, combinations of zinc selenide, zinc sulfide, and KRS5 (thallium bromo-iodide) appear promising, where the higher refractive index of these materials enables lower spherical aberration for surfaces of equivalent power~\cite{Smith2005}.  Further still, a change in laser and monitoring wavelengths could enable use of silicon and germanium as optical materials and this high-index combination is recognized as effective for achromatic lens designs at these wavelengths~\cite{Smith2005}.  The final option is a mirror-based system, which is inherently free of chromatic aberration~\cite{Kasunic2011, Smith2005}, perhaps following the off-axis, three-mirror anastigmat designs of Korsch~\cite{Korsch1991}. 

Next, the process signatures and alarm criteria extracted here represent only a handful of the ways in which the video data can be reduced to process signatures or alarms.  It may be that a combination of process signatures features higher predictive power than any individual process signature.  This is particularly motivated by a qualitative reading of Fig.~\ref{fig:Fig5}c, where the different signatures appear sensitive to different aspects of the process (i.e., the process signatures are linearly independent).  There is also considerable room for investigating alternative process signatures.  Inspiration may be found, for example, using the clustering and principle component analysis techniques applied to detecting spatter events in~\cite{Grasso2016, Yan2022} and the spectral graph theory of~\cite{Montazeri2018}.  Likewise, machine learning techniques such as neural networks have already shown promise~\cite{Okaro2019, Jayasinghe2022, Dewinton2021}.  Finally, the process input is well-defined via the scan file, and these data may be used to help interpret the camera data.  Inspiration for this this direction is found in nonlinear system identification techniques, such as the NARMAX model (see, e.g.~\cite{Billings2013} and the many references therein).  Other nonlinear analysis methods are also promising, including our preliminary investigation of recurrence analysis in~\cite{Penny2024_Thesis}.

Residual stress and microstructure are also affected by the complex and spatially varying time-temperature history inherent to LPBF~\cite{Mercelis2006, Bartlet2019,Acharya2017,Kohnen2019,Leicht2020}; accordingly, it is expected that a correlation may be found from the present process signatures to these material attributes as well.  Neutron diffraction is sometimes used map residual stress tensors in this context~\cite{Wu2014,Cornwell2018,ORNL2023} among other options for obtaining ground truth including synchrotron X-ray diffraction, hole drilling, and sectioning techniques~\cite{Rossini2012}.  Likewise, microstructure data can be obtained with a variety of methods, including light microscopy after etching, microindentation, and electron imaging techniques like scanning electron microscopy (SEM) and electron backscatter diffraction (EBSD)~\cite{Smallman2007}. 

Finally, while not investigated in this work, there is interest in using synchronized motion of two lasers to manipulate temperature profiles about the material being fused.  This can be done to beneficially alter the build rate, quality, microstructure, and residual stress (see, e.g.,~\cite{Heeling2018, Wong2019, Tsai2019, Cao2021, Chen2021, Li2021, Wei2021, Zhang2022, Promoppatum2022}).  The difficulty to achieving these aims with conventional LPBF equipment is that the scan area of an f-theta lens can be comparable to the dimensions of the lens itself.  In this case, the achievable overlapping area for two lasers is only a narrow stripe as noted directly in~\cite{Heeling2018} and, while not explicitly discussed, is evidenced in~\cite{Li2021} and~\cite{Wei2021}.  An alternate implementation of ADM can enable coordinated multi-laser processing to be applied across the full extent of the build area.

\section{Acknowledgements}

This study was funded by Honeywell Federal Manufacturing \& Technologies (FM\&T) and by a MathWorks MIT Mechanical Engineering Fellowship (to R.W.P.). The funding bodies played no role in study design, data collection, analysis and interpretation of data, or the writing of this manuscript. We thank Dan Gilbert, Paul Carson, and Joe Wight of the MIT LMP machine shop for facilitating assembly of the instruments described herein.

%\section*{References}
\bibliographystyle{elsarticle-num}
\bibliography{refs09_27_2021.bib}

\section{Author Contributions}

R.W.P. and A.J.H. conceptualized ADM, designed the experiments, and wrote the manuscript. R.W.P performed the design of the ADM lens, performed the experiments, and analyzed the results. Z.K. contributed to implementation of the LPBF testbed including development of experimental protocols and identification of suitable process parameters.  A.J.H. supervised the project, including guiding data analysis, obtaining funding, and securing project resources. All authors read and approved the final manuscript.

\beginsupplement
\pagebreak

\section{LPBF Testbed}
\label{sec:SupTestbed}

\begin{figure*}[htbp]
\begin{center}
	{
	\includegraphics[trim = {1in 1in 1in 2.75in}, clip, scale=1, keepaspectratio=true]{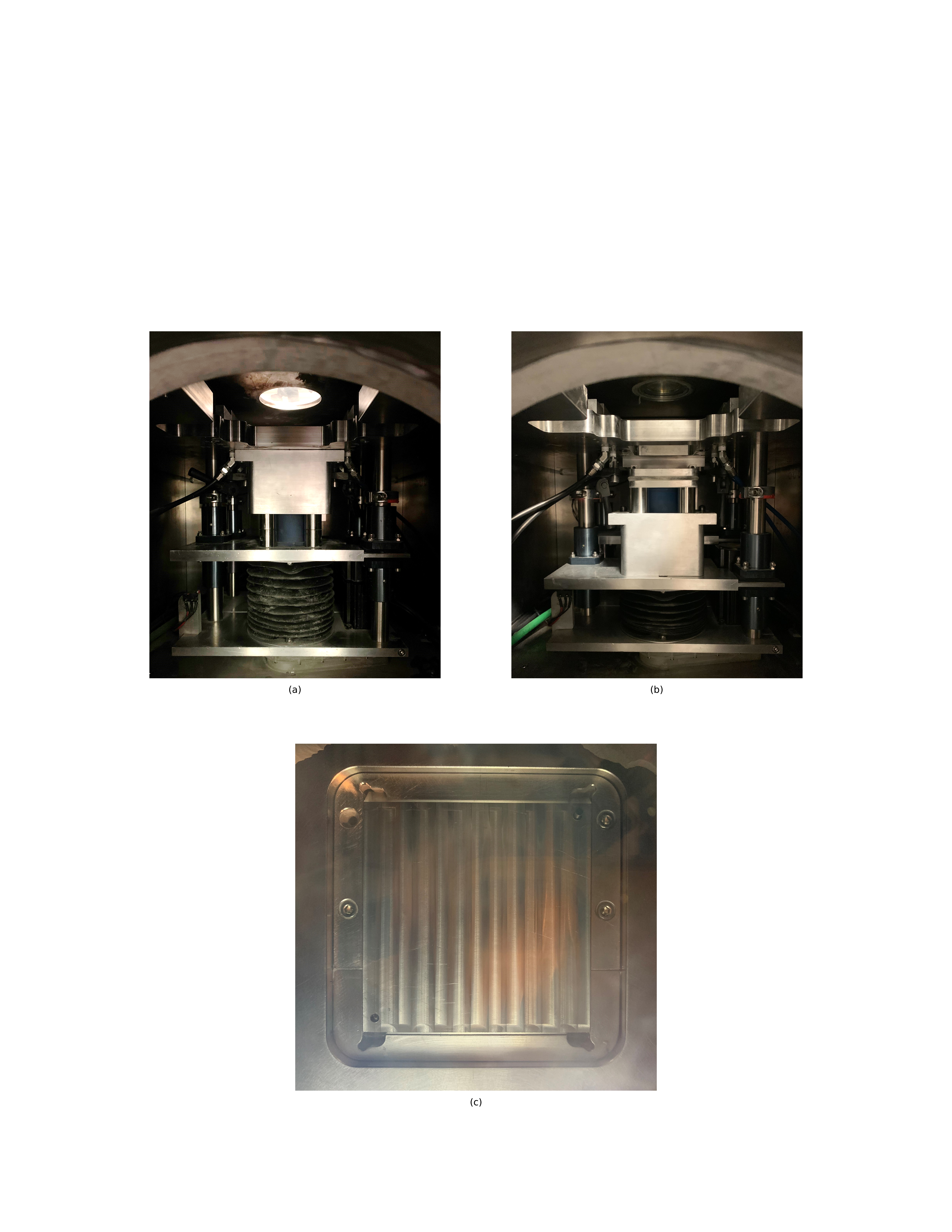}
	}
\end{center}
\caption{Design features of the LPBF testbed that enable print removal from the front of the machine.  (a) Typical printing configuration.  (b)  Printer interior after part removal and ready to receive a build plate for the next print.  The build piston bore has been detached and slid downwards as compared to (a), exposing the top of the build piston.  (c) Top view of an installed build plate, showing the clamping mechanism securing the build plate to the build piston.}
\label{fig:Fig8}
\end{figure*}

Figure~\ref{fig:Fig8} details the unique mechanical features that allow for build plate installation and part removal from the front of the LPBF testbed.  Specifically, Fig.~\ref{fig:Fig8}b shows a view though a front hatch, where the printer is configured to receive a build plate.  Access to the top of the build piston is provided by detaching the piston bore from the printer structure and sliding it downwards.   The build plate is then installed from the front, and is secured with the retainer mechanism shown from the top in Fig.~\ref{fig:Fig8}c.  The piston bore is then slid upwards and clamped in position, shown in Fig.~\ref{fig:Fig8}a, as the final step before printing.  Removal of the component after printing simply reverses these steps.

\section{ADM Lens}
\label{sec:SupADM}

\subsection{Merit Function Operands}

The merit function has operands operands for:

\begin{enumerate*}
	\item Directing the effective focal length of both configurations, evaluated in both the X and Y directions, to a target value of $125$~mm.
    \item For the laser spots corresponding to each field:
    \begin{enumerate*}
    	\item Directing the encircled energy within a $32.5$~\textmu m radius circle to a target value of 86\% (or that the D86 is $65$~\textmu m).
    	\item Directing the encircled energies within the stripes $\pm 32.5$~\textmu m along each of the X and Y axis to be equal (or that the spots are roughly symmetric).
    	\item Maintaining the encircled energy within a $25$~\textmu m radius circle to a target value of 50\% (or that the spots are roughly gaussian).
    \end{enumerate*}
    \item Minimizing the polychromatic RMS spot size for each field of the imaging path.
    \item For the physical design:
    \begin{enumerate*}
    	\item Directing the airspace between element edges to be at least $0$~mm (or that the lens element edges cannot overlap).
    	\item Directing the airspace between element centers to be at least $0$~mm (or that the lens element centers cannot overlap).
    	\item Directing the airspace between element centers to be less than $45$~mm (excepting an unconstrained airgap between each lens group).
    	\item Directing that the element edge thicknesses be at least $4.5$~mm
    	\item Directing that the element center thicknesses be at least $4.5$~mm
    	\item Directing that the element center thicknesses be less than $45$~mm
    \end{enumerate*}
\end{enumerate*}

\subsection{Tolerance Analysis}

A tolerance analysis is performed for each optical path in the final ADM lens design independently, using the original, forward model to consider the laser path and the second, reversed model to consider the imaging path.  Only the operands from the merit function relevant to the specific path are retained for assessment of performance degradation.  As the split merit functions have unmixed units, their interpretation is also more straightforward.  Tolerance operands specify:
\begin{itemize*}
	\item Powered element radius: $\pm 0.2$\%,
	\item Plano surface curvature: $3$~waves at $633$~nm,
	\item Center thickness: $\pm 100$~\textmu m,
	\item Element decentration (X and Y): $\pm 50$~\textmu m,
	\item Element surface total indicated runout (TIR, X and Y): $\pm 0.025$~mm, and
	\item Surface irregularity: $\pm 1$~wave.
\end{itemize*}

For the laser path, the merit operands are all GENF (geometric encircled energy fraction) that returns the fraction of energy encircled at a given distance.  For the nominal design, the average deviation in encircled energy at the targeted diameters is about 6.3\% (i.e. about 80\% or 92\% of the laser light could be going through a $65$~\textmu m diameter instead of the 86\% target.)  A Monte Carlo analysis of $200$ systems, assuming a pessimistic parabolic statistical distribution, suggests that the mean as-built system has an average deviation of about $8$\%, with a standard deviation of $5.7$\%.  Sensitivity analysis shows the worst offenders to be the TIR and decenters on the elements described by Surfaces~7 and~8, and~9 and~10 in Table~\ref{table:BroadbandPrescription}, respectively.  This is not a surprise, as these elements have some of the tightest radii in the design.  Nevertheless, despite the worst-case in the Monte Carlo analysis indicating $30$\% deviations from the specified encircled energy values, the present tolerances are extremely likely to yield a serviceable assembly because adjustments to laser and scan speed can partially compensate for a deviation in spot size.

All merit operands for the imaging path evaluate the RMS (root-mean-square) spot radius for each field at the image plane: the nominal average is $2.33$~\textmu m.  Using the same parameters for a Monte Carlo tolerance analysis shows the mean spot size is expected to be $2.9$~\textmu m with a $0.6$~\textmu m standard deviation across the ensemble.  The worst case found has a spot size of $5.5$~\textmu m, which, while a substantial hit against the theoretical performance, remains reasonably small as compared to the $20$~\textmu m pixel pitch.  Again, the most sensitive elements are indicated by Surfaces~7 and~8, and~9 and~10 in Table~\ref{table:BroadbandPrescription} in the sensitivity analysis.  A root-sum-square (RSS) analysis of the sensitivity study is in agreement with the Monte Carlo approach, suggesting a spot size of $2.7$~\textmu m.  Imaging is arguably the more vital requirement to achieve and this analysis shows that the present tolerances provide fully-sufficient performance.

\subsection{Optical Materials and Coatings}

Calcium fluoride and IR-grade fused silica are specified in the nominal ADM lens design.  Unfortunately, due to sourcing difficulty at the time of manufacture, the fused silica used in the ADM lens is UV-Vis grade (Corning 7980) and not IR-grade material.  The only practical difference lies in hydrogen contamination present in the UV-Vis grade material, which causes an absorption feature at $2.73$~\textmu m, and significant overtones at $2.24$ and $1.37$~\textmu m~\cite{Moore2022}.  For the present experiments, these absorption features are of little consequence.  An anti-reflection (AR) coating is specified on all optical surfaces to reduce Fresnel reflections.  This takes priority at the laser wavelength, where reflection at each surface must be less than 0.25\%, primarily to prevent scattered laser light from heating the ADM lens assembly.  A secondary specification is a broadband AR coating from~$1.2$ to~$2.4$~\textmu m, although it is something of a compromise with the laser specification.  Inspection data show $ R = 0.0907$ at $1070$~nm and $R = 0.8046$\% on average from $1.2$ to~$2.4$~\textmu m for the fused silica coating and corresponding values of $0.1125$ and $0.8288$\% for the CaF$_2$ coating.

\subsection{Additional Mechanical Details}

A lens-tube approach, wherein all the optical elements are radially constrained within a single tube in two groups, is used to avoid the tolerance stack of a multi-part assembly.  Conical lens seats locate the bottom-most optic of each group and all the optics are stacked from the top.  The optics are spaced with spacing rings made of 360~brass.  These components are also designed to minimize contact stress by tangentially contacting the surfaces of the optics (i.e., these lens seats are also conical).  Fine-threaded brass retainers are used to secure each group.

An adjustable kinematic coupling assembly is used to locate the ADM lens tube on top of the LPBF testbed environmental enclosure.  The lens tube itself is attached to an interfacing component, which also features three Thorlabs M6x0.25 fine adjustment screws.  Three feet form the other half of the coupling, which are permanently affixed to the top of the LPBF testbed enclosure.  Each of these feet comprise two horizontal M6x36 dowel pins.  As such, the position of the ADM lens is defined by 6 point contacts, where each of the adjustment screw ball ends contacts two of the M6 pins.  The coupling is preloaded gravitationally, i.e., the ADM lens simply stands on the feet without additional fixturing.  Focusing the ADM lens on the build platform is preformed with these fine adjustment screws using the same sensor and procedure used to qualify the laser focus in \S~\ref{sec:MethodsQualification}.

\section{Additional Detection Thresholds}
\label{sec:DetectionThresholds}

\begin{figure*}[htbp]
\begin{center}
	{
	\includegraphics[trim = {1.2in 3.25in 1.2in 3.5in}, clip, scale=1, keepaspectratio=true]{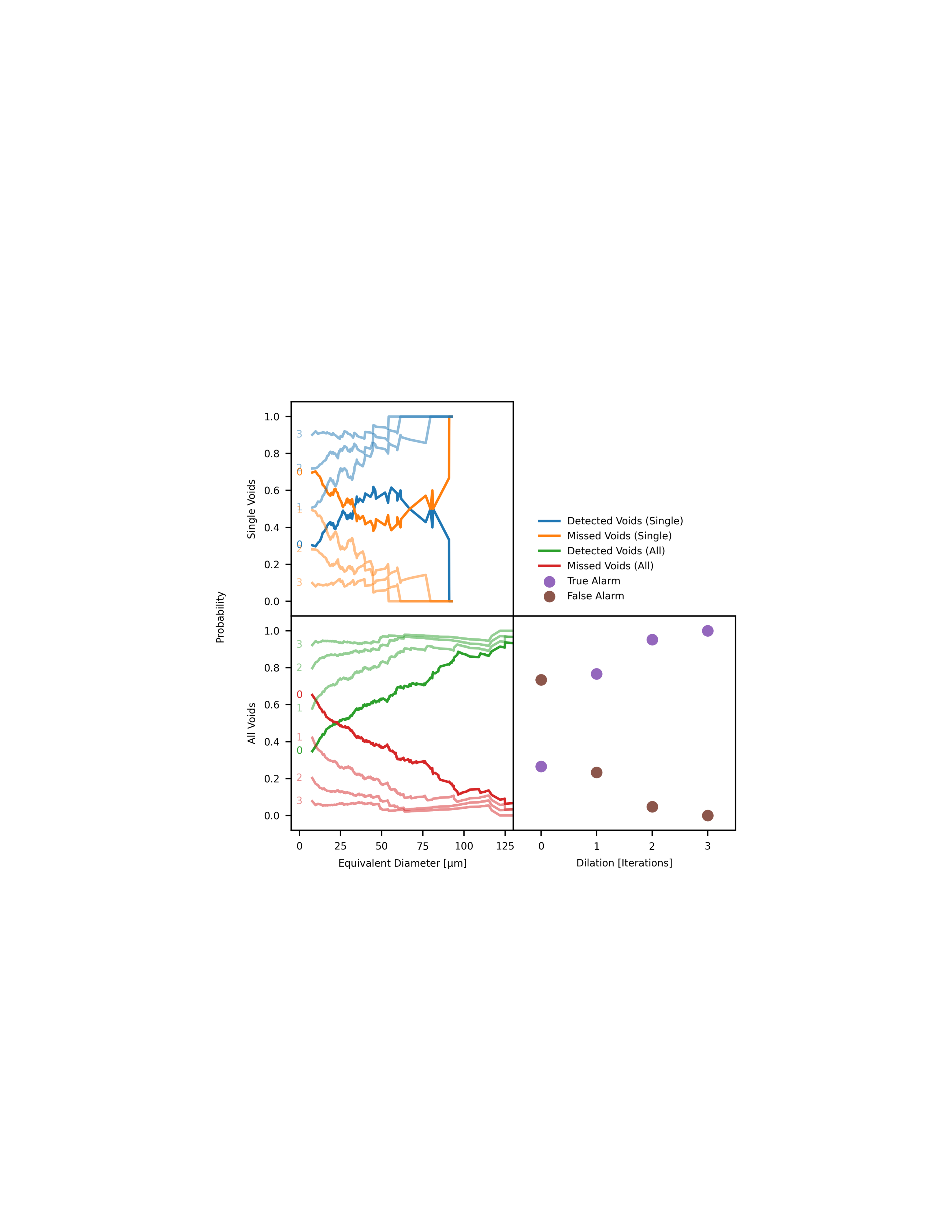}
	}
\end{center}
\vspace{-11pt}
\caption{Detection probabilities using maximum radiance as a high-threshold ($\geq 8000 \, \mathrm{counts}$) process signature.  Panels directly correspond to those in Fig.~\ref{fig:Fig6}a.}
\label{fig:Fig9}
\end{figure*}

\begin{figure*}[htbp]
\begin{center}
	{
	\includegraphics[trim = {1in 3.25in 1in 3.5in}, clip, scale=1, keepaspectratio=true]{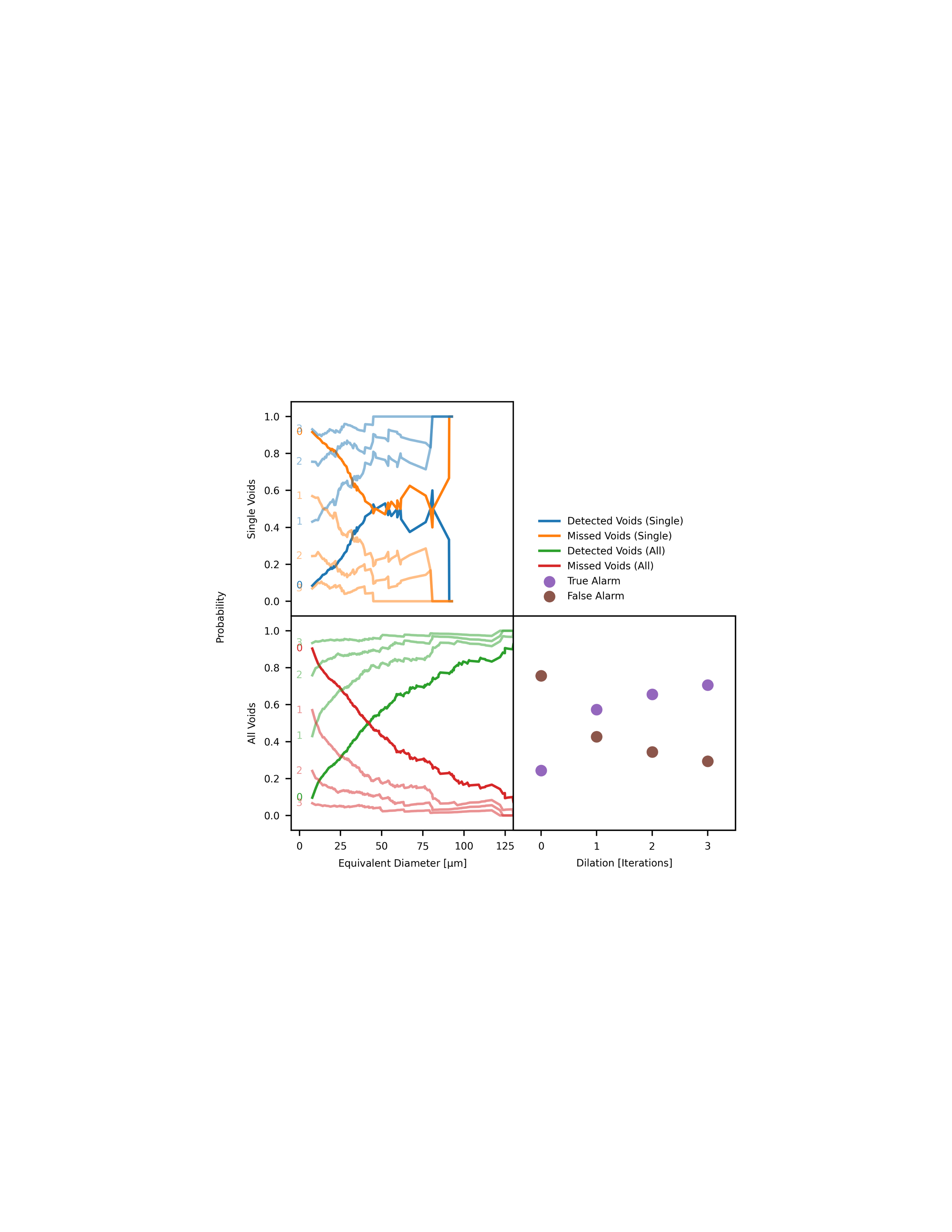}
	}
\end{center}
\caption{Detection probabilities using meltpool size at last melting event as a high-threshold ($\geq 15 \, \mathrm{px}$) process signature.  Panels directly correspond to those in Fig.~\ref{fig:Fig6}a.}
\label{fig:Fig10}
\end{figure*}

\end{document}